\documentclass[useAMS,usenatbib]{mn2e}
\usepackage{amssymb,amsmath,epsfig,times, natbib}
\title[Hot gas galaxy halo]{A deep \emph{Chandra} observation of the hot gaseous halo around a rare, extremely massive and relativistic jet launching spiral galaxy}
\author[S. A. Walker et al.]{S. A. Walker,$^1$\thanks{Email: 
    swalker@ast.cam.ac.uk} J. Bagchi$^2$ and A. C. Fabian$^1$  \\
  $^1$Institute of Astronomy, Madingley Road, Cambridge CB3 0HA \\
  $^2$The Inter-University Centre for Astronomy and Astrophysics (IUCAA), Pune University Campus, Post
Bag 4, Pune 411007, India \\
  \\
    \\
   \\
   \\
}
\date{}

\begin{document}

\maketitle

\begin{abstract}
We present a deep Chandra observation of the extremely massive spiral galaxy 2MASX J23453268-0449256, the first X-ray observation of this very rare system which features the largest known relativistic jets from a spiral galaxy. We detect extended X-ray emission from the hot halo surrounding the galaxy, reaching out to 80 kpc in radius. The hot halo is elongated along the plane of the spiral galaxy, and one possibility is that the powerful relativistic jets have disrupted the hot halo gas located perpendicular to the disk. Our calculations indicate that it is energetically feasible that the AGN feedback in this system could have uplifted or completely expelled a significant fraction of the gas in the 20-80 kpc radial range. We also detect extended emission which appears to be associated with the inner and outer southern radio lobes, and is possibly the result of inverse Compton emission. Using the observed X-ray and radio luminosity of the central AGN, the fundamental plane of Gultekin et al. predicts a black hole mass of 5$\times10^{8}$ M$_{\odot}$, with a range of $1\times10^{8}$ - $3\times10^{9}$ M$_{\odot}$ when the scatter in the fundamental plane relation is taken into account. This is consistent with the possibility that an exceptionally massive ($>10^{9}$ M$_{\odot}$) black hole lies at the centre of this galaxy, as suggested by the $M_{BH}-{\sigma}$ scaling relation, but a tighter constraint cannot be made.
 
\end{abstract}

\begin{keywords}
galaxies: individual (2MASX J23453268-0449256) - galaxies: ISM - galaxies: spiral - X-rays: galaxies - X-rays:
general - X-rays: ISM
\end{keywords}

\section{Introduction}
Galaxy formation models have made the important prediction that galaxies are surrounded by hot gaseous halos since \citet{White1978}. These hot halos are predicted to form through the accretion of matter
onto the dark matter halo, with shocks raising the baryons to the virial
temperature (\citealt{White1991}; \citealt{Benson2010}). It is predicted that the baryonic mass in these halos is comparable to or even exceeds the baryonic mass of the galaxies themselves, depending on the levels of pre-heating, galactic feedback heating, and cooling rates that are assumed (\citealt{Sommer-Larsen2006}; \citealt{Fukugita2006}). 

Observations of nearby galaxies have found that they are missing most of
their baryons (e.g., Hoekstra et al. 2005; \citealt{Heymans2006};
\citealt{Mandelbaum2006}; \citealt{Gavazzi2007}; \citealt{Jiang2007}; \citealt{Bregman2007}) compared to the expected mean cosmic
baryon fraction determined by WMAP (f$_{b}$ = 0.171 $\pm$ 0.009; \citealt{Dunkley2009}). In other galaxies, a variety of methods has confirmed that the baryonic mass is much lower than expected (e.g., \citealt{Hoekstra2005};
\citealt{McGaugh2005}). It is possible that these hot gas halos contain the `missing baryons' from these galaxies, bringing the total baryon budget of the galaxies into agreement with expectations based on the mean cosmic baryon fraction.

Extensive observations of hot halos around early-type galaxies in soft X-rays (0.5-2 keV) have already been made (\citealt{Forman1985}; \citealt{OSullivan2001}; \citealt{Mulchaey2010}). 
However, as these galaxies have become elliptical, coronal gas can be produced through the merging process and associated star formation (\citealt{Read1998}). This makes it very difficult to accurately connect these halos with the galaxy formation process. These ellipticals are also commonly located in the centres of groups and clusters, and it can be challenging to separate the galaxy halo emission from the surrounding intragroup medium.

Greater insight can be achieved by studying the hot halos around disk galaxies, which should
be much more direct tracers of the galaxy formation process since they have not undergone major merging. Since the hot halo mass and X-ray emission scales with galaxy mass, it is necessary to observe the most massive spiral galaxies to achieve a detectable X-ray signal.  

Detecting soft X-ray haloes beyond the optical radii of spiral galaxies has proven to be very challenging, with only upper limits being derived in early studies with ROSAT (\citealt{Benson2000}) and Chandra (\citealt{Rasmussen2009}). Detections of emission extending a few kpc above the disks of spirals have been made, but these are the result of star formation in the galaxies, and are likely due to galactic fountains (e.g. \citealt{Strickland2004}, \citealt{Li2006}, \citealt{Tullmann2006}). 

It is only recently that a breakthrough in the observation of hot haloes around disk galaxies has been made, using Chandra and XMM-Newton to detect the hot halos in a handful of extremely massive, fast rotating spiral galaxies: NGC 1961 (\citealt{Anderson2011}), UGC 12591 (\citealt{Dai2012}), NGC 6753 (\citealt{Bogdan2013a}) and NGC 266 (\citealt{Bogdan2013b}). For these galaxies, extended soft X-ray emission has been detected out to $\sim$80 kpc, far outside their optical radii. When the best fitting surface brightness models for these hot haloes were extrapolated out their virial radii, the total baryon fractions inferred were, however, still below the cosmic baryon fraction for all of these galaxies. Possible explanations for the missing baryons include AGN and supernova feedback expelling gas from the galactic potential well, and the possibility that preheating of the gas has prevented it from falling into the potential well as the galaxy formed. 

Here we present a deep, 98.7ks, Chandra observation of the extremely massive, rapidly rotating  ($V_{rot}$=429$\pm$30 km s$^{-1}$), relativistic jet launching, spiral galaxy 2MASX J23453268-0449256 (hereafter J2345-0449) to detect its hot halo and measure its mass content. Our Chandra observation is the first time this system has been detected in X-rays. 

We use a standard $\Lambda$CDM cosmology with $H_{0}=70$  km s$^{-1}$
Mpc$^{-1}$, $\Omega_{M}=0.3$, $\Omega_{\Lambda}$=0.7. For the redshift of J2345-0449, $z=0.0755$, the luminosity distance is 342 Mpc and the angular scale is 85.9 kpc/arcmin. All errors unless
otherwise stated are at the 1 $\sigma$ level. 

\section{Previous Observations of J2345-0449}

The radio and optical properties of this extremely rare and exciting galaxy have been studied in \citet{Bagchi2014}. \citet{Bagchi2014} studied the kinematics of the optical Balmer H$_{\alpha}$ line observed with the IUCAA Girawali Observatory (IGO) 2m telescope, and found an extremely large rotation speed, V$_{rot}$ = 371/sin($i$) = 429 ($\pm$30) km s$^{-1}$, in the asymptotic flat region at r $\geq$ 10 kpc from the galactic center. This makes J2345-0449 one of the most massive known spiral galaxies. Another striking result obtained in \citet{Bagchi2014} is the exceptionally large stellar velocity dispersion $\sigma$ = 326($\pm$59) km/s measured within the central (3 arcsec, or 4.3 kpc) region, higher than for the majority of bulge-less
disks on such a spatial scale. This suggests a huge central concentration of mass, including a
SMBH, and the lower limit from the optical data is $>$ 2$\times$10$^{8}$M$_{\odot}$.

\citet{Bagchi2014} found that J2345-0449 features a pseudobulge (with a mass of 1.07($\pm$0.14) $\times10^{11}$ M$_{\odot}$), whose small Sersic index and scale radius, and low bulge-to-total light ratio make it similar to pseudobulges observed in other systems (e.g. \citealt{Gadotti2009}). This is discussed in detail in \citet{Bagchi2014}.

\begin{figure*}
  \begin{center}
    \leavevmode
    \hbox{
      \epsfig{figure=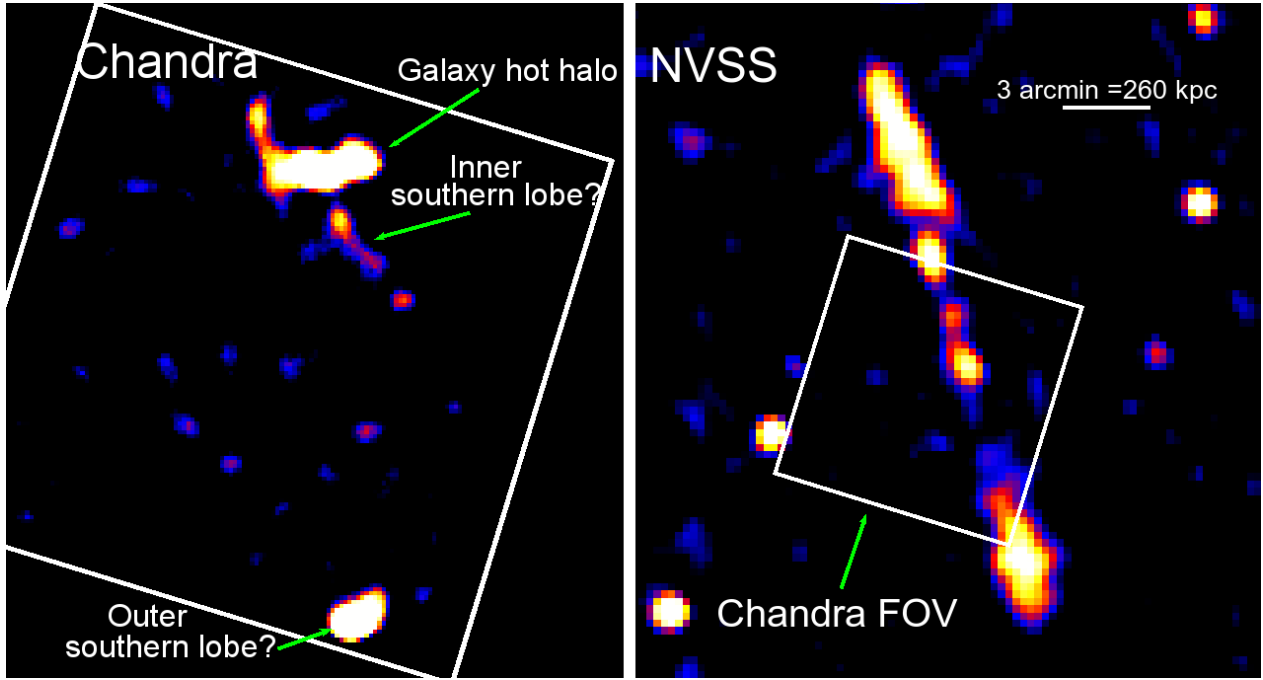,
        width=0.95\linewidth}
        
            }
    
      \caption{\emph{Left}:Soft band (0.5-2.0 keV) exposure corrected image of the whole Chandra field of view. Point sources detected with \textsc{wavdetect} have been removed, the image binned by a factor of 8 and the image smoothed with a Gaussian kernel with a radius of 6 pixels. X-ray features coinciding with the inner and outer southern lobes are highlighted. \emph{Right}: NVSS image of the radio emission, showing the inner and outer lobes, with the Chandra field of view shown in the left panel overplotted as the white square.}
      \label{NVSScompare}
  \end{center}
\end{figure*}

\citet{Bagchi2014} also found that this galaxy is currently ejecting a collimated pair of relativistic jets out to $\sim$ 415kpc. There are also larger (the largest yet seen) and more diffuse older radio lobes from previous jet activity reaching over $\sim$ 1.6 Mpc which lack prominent hot spots and are no longer being energized by the jets. Since radio sources appear to require diffuse hot gas to act as a working surface for making the lobes, extended X-ray emission is expected in this system.

The close alignment of the inner and outer of the radio lobe pairs indicates that the spin axis of the central black hole has remained relatively stable over the timescale of $\sim10^{8}$ yr between jet triggering episodes (\citealt{Bagchi2014}). \citet{Bagchi2014} conclude that the galaxy and central black hole have evolved together quietly, with no major merging activity. \citet{Bagchi2014} found that this quiet evolution is likely due to the disklike nature of the pseudobulge, the absence of any tidal debris (such as plumes, stellar streams or shells), and the fact that the galaxy has a stable rotationally supported stellar disk featuring well formed spiral arms.  The galaxy is also located in a relatively isolated galactic environment, with no nearby luminous galaxies, and it is not in the centre of a group or galaxy cluster.

It is extremely rare for a massive spiral galaxy to eject relativistic jets, as they are nearly always launched from the nuclei of bulge dominated ellipticals and not flat spirals. It is clear that J2345-0449 is an extremely rare system whose properties challenge the standard paradigm for the formation of relativistic jets in AGN. The fact that the jets are aligned almost exactly perpendicular to the plane of the disk would also suggest that the central SMBH has a high spin. The detailed X-ray observations we present here are necessary to fully understand this system and complement the excellent GMRT and VLA radio data.

To determine the expected virial mass ($M_{200}$) of the galaxy, we follow the approach of \citet{Bogdan2013a} and use the baryonic Tully-Fisher relation (\citealt{McGaugh2005}) for the cold dark matter cosmogony, $M_{200} \propto V_{\rm max}^{3.23}$, which relates the virial mass to the maximum rotational velocity ($V_{\rm max}$=430 km s$^{-1}$). As in \citet{Bagchi2014}, we find that, scaling from the virial mass found for NGC 1961 in \citet{Bogdan2013a}, we obtain $M_{200}=1.05 \times 10^{13}$ M$_{\odot}$ and $r_{200}$=450 kpc.


\begin{figure*}
  \begin{center}
    \leavevmode
    \hbox{
      \epsfig{figure=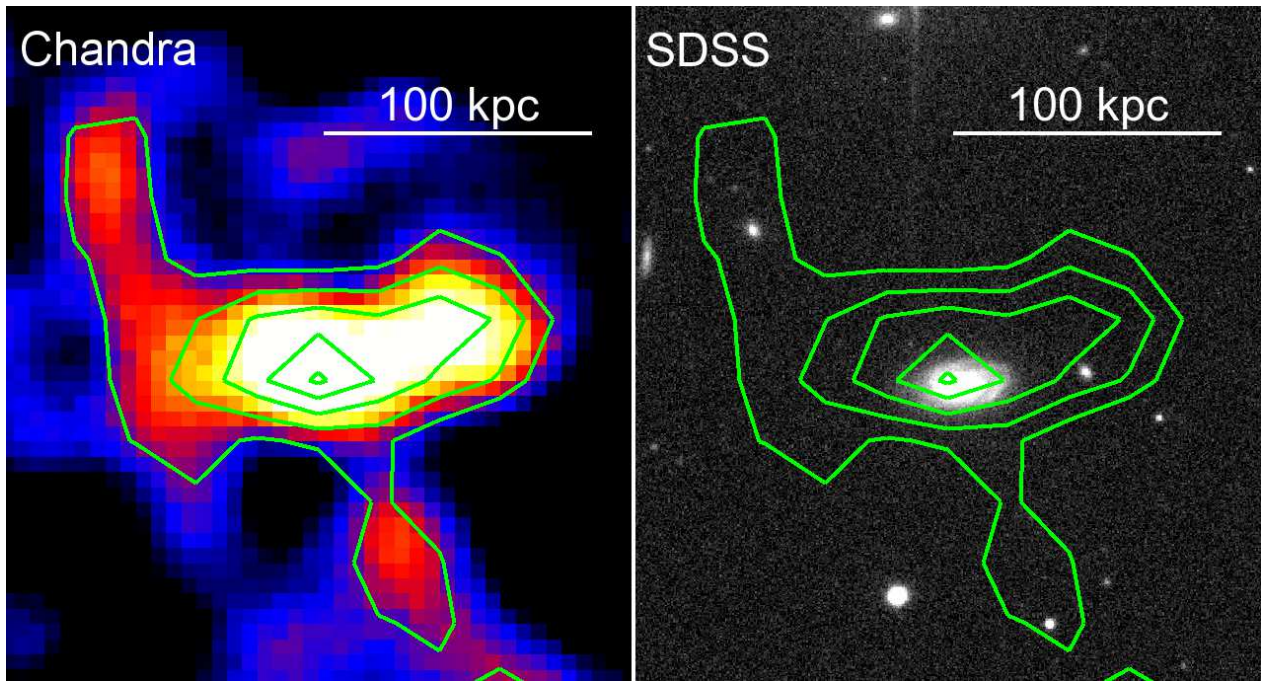,
        width=0.95\linewidth}
            }
    
      \caption{\emph{Left}: Soft band (0.5-2.0 keV) exposure corrected Chandra image which is the same as the left panel of Fig. \ref{NVSScompare} but zoomed in on the galaxy. \emph{Right}: SDSS image of the galaxy, with the X-ray contours overplotted in green corresponding to surface brightnesses of (from the centre outwards) 2, 1, 0.8, 0.7 and 0.6$\times10^{-6}$ cts/s/arcsec$^{2}$, chosen to be relatively evenly distributed spatially. Both images have had their coordinate systems matched.}
      \label{SDSScompare}
  \end{center}
\end{figure*}

\begin{figure*}
  \begin{center}
    \leavevmode
    \hbox{
      \epsfig{figure=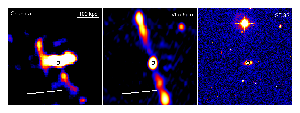,
        width=0.95\linewidth}
            }
    
      \caption{\emph{Left}: Soft band (0.5-2.0 keV) Chandra image, which is the same as the left panel of Fig. \ref{NVSScompare} but zoomed in on the galaxy and the X-ray feature coincident with the inner southern radio lobe highlighted by the white arrow in both panels. \emph{Centre}: VLA 6cm radio image of the inner lobes \emph{Right}: SDSS image of the galaxy showing the surrounding environment. All three panels have their coordinates matched, and the black circle centred on the galaxy is in the same location in each panel. }
      \label{southernlobefig}
  \end{center}
\end{figure*}

\section{Observations, Data Reduction and Images}

Chandra observed J2345-0449 for 98.7ks between 2014-08-23 and 2014-08-24 (PI S. A. Walker), using the ACIS-S detector owing to its higher effective area than ACIS-I. As shown in Fig. \ref{NVSScompare}, the Chandra observation covers the galaxy, the southern inner radio lobe and part of the southern outer radio lobe.

The Chandra observation was reduced using CIAO version 4.6, using the latest calibrations. The events file was reprocessed using the task \textsc{chandra\_repro}. We examined lightcurves in the 0.5-7.0 keV to identify any periods of flaring that needed to be removed. We used the routine \textsc{lc\_sigma\_clip} to identify periods where the count rate differed from the mean by more than 2 $\sigma$. The count rate for the observation is very stable, and only brief periods needed to be removed. After all of this filtering, the resulting clean exposure time was 97.8ks.

The task \textsc{flux\_obs} was used to extract images in a soft band (0.5-2.0 keV), which was chosen to maximise the signal to noise of the detection of the hot halo surrounding the galaxy, and a hard band (2.0-6.0 keV), which is used later to determine the contribution of the extended X-ray emission from Low Mass X-ray Binaries (LMXBs). An image was also extracted in a broad, 0.5-7.0 keV band to enable point source detection and removal. Point sources were identified from the broad band image using \textsc{wavdetect}, using a range of wavelet radii between 1 and 16 pixels to ensure that all point sources were removed, and using a PSF (point spread function) map generated using \textsc{mkpsfmap}. 

We first remove the particle background from the image. This is achieved by using stowed background events files. These are observations taken with the ACIS-S detector moved out of the focal plane, such that the events detected all originate from just the particle background. The ACIS-S stowed events file nearest in time to our observation was reprojected into the same sky coordinates as our observation, and rescaled to have the same exposure time. The stowed events file was then rescaled so that the count rate in the 9.5-12.0 keV band matched that of our observation. In this 9.5-12.0 keV band, the effective area of the Chandra telescope is zero, so the only events in our observation in this band originate from the particle background. We then extracted images from the rescaled and reprojected stowed events file in the same bands as the images we extracted from our observation: in soft (0.5-2.0 keV), hard (2.0-6.0 keV) and broad (0.5-7.0 keV) bands. 

Since the particle background is not vignetted, these particle background images are subtracted from the corresponding images from our observation before any exposure map correction is performed. This removes the particle background component from our observations. \citet{Hickox2006} have shown that the particle background spectrum for Chandra is highly stable, remaining constant to within 2 percent even when the overall flux changes. We explored the systematic effect of a conservative $\pm$ 4 percent variation in the particle background level and spectral shape, and found this to have no impact on the X-ray detection of the galaxy. 

Exposure maps in the soft, hard and broad bands were created by the task \textsc{flux\_obs} using as the spectral weight the spectrum of the the entire ACIS-S3 chip. We also experimented with using monochromatic exposure maps at energy 1.2 keV and 3.4 keV for the soft and hard band images respectively, and found that the effect on the 
resulting exposure corrected images was negligible. The particle background subtracted images were then divided by the corresponding exposure maps to create exposure corrected soft, hard and broad band images.

The remaining X-ray background originates from our galaxy (the Local Hot Bubble and the Milky Way halo) and also from unresolved point sources (extragalactic AGN). The long exposure time allows point sources to be removed down to a very low threshold flux (5$\times 10^{-16}$ erg cm$^{-2}$ s$^{-1}$), and we find that the systematic errors arising through fluctuations in the expected level of the cosmic X-ray background also have an insignificant affect on our measurements. We used local background regions away from the galaxy and jets to measure the remaining X-ray background level. The local X-ray background is very uniform, with very low spatial variations of $\pm$ 3 percent on the scale of the region used to extract the counts from the hot halo of the galaxy. We investigated the systematic effects of varying the galactic background by its observed variance across the field, and found this had a 
negligible affect on the detection significance of the galaxy, and on the surface brightness profiles we extract and describe later in section \ref{sb_prof}.

We detect 100$\pm14$ counts from the hot halo, and 70$\pm10$ counts from the central AGN in the 0.5-2.0 keV band. These errors are the sum of the statistical errors and the systematic errors obtained by varying the particle background by $\pm$4 percent, and varying the X-ray background by its observed variance of $\pm$3 percent over the background regions. The region used to extract the hot halo counts is shown in Fig. \ref{haloregion} in Appendix \ref{AppendixA}. It excludes the central 0.05$r_{200}$=15.5 kpc shown by the dashed circle in this figure. The left panel of Fig. \ref{NVSScompare} shows the soft band, point source and particle background subtracted, and exposure corrected image which has been binned by a factor of 8 and then smoothed using a Gaussian kernel of radius 6 pixels. The central AGN has also been removed from this image. There is clear extended emission around the galaxy J2345-0449.  In Fig. \ref{SDSScompare} we show a zoom in of the soft X-ray galaxy emission, compared to the SDSS image of the galaxy. The X-ray emission appears elongated along the plane of the galaxy disk, but is much larger in scale, with a diameter of around 160 kpc. This far exceeds the optical radius of the galaxy (25 kpc), showing that we are detecting emission from the hot halo.

The bolometric X-ray luminosity of the detected parts of the extended halo in the radial range 0.05-0.15$r_{200}$ is $8.0\pm2.2 \times 10^{40}$ erg s$^{-1}$. This is in reasonable agreement with the total bolometric X-ray luminosities obtained in \citet{Bogdan2013a} for NGC 1961 and NGC 6753 in the same radial range, which are $5.8\pm1.7 \times 10^{40}$ erg s$^{-1}$ and $6.3\pm0.9 \times 10^{40}$ erg s$^{-1}$ respectively.

The shape of the extended X-ray halo emission is incompatible with a galactic fountain origin due to star formation activity. Galactic fountains lead to extended emission above and below the disk (where we actually observe a lack of extended emission) and do not produce extended X-ray emission outside the optical radius of the disk (e.g. \citealt{Strickland2004}). 

There are also suggestions of extended emission from the inner southern radio lobe and part of the outer southern lobe covered by the observation. No galaxies are coincident with this southern extended emission, so it is possible that we are seeing inverse Compton emission from the inner and outer southern lobes. A zoom in comparison of the inner southern lobe X-ray and radio features is shown in Fig. \ref{southernlobefig}. We detect 22$\pm$7 counts from the inner lobe (a 3 sigma detection), and 60$\pm$10 counts from the small part of the outer lobe (just 10 percent) we cover. In both cases the systematic uncertainties in the background components have been propagated into the error budget.

\section{Surface Brightness profiles}

\label{sb_prof}
The surface brightness around the galaxy appears asymmetric, with the bulk of the emission occurring in the direction of the plane of the galaxy. It is possible that the powerful radio jets in this system have disturbed the hot halo perpendicular to the plane, removing gas from it. To maximise the signal to noise of the surface brightness profile of the hot halo, we initially consider only the directions away from the jets

\begin{figure}
  \begin{center}
    \leavevmode
    \vbox{
      \epsfig{figure=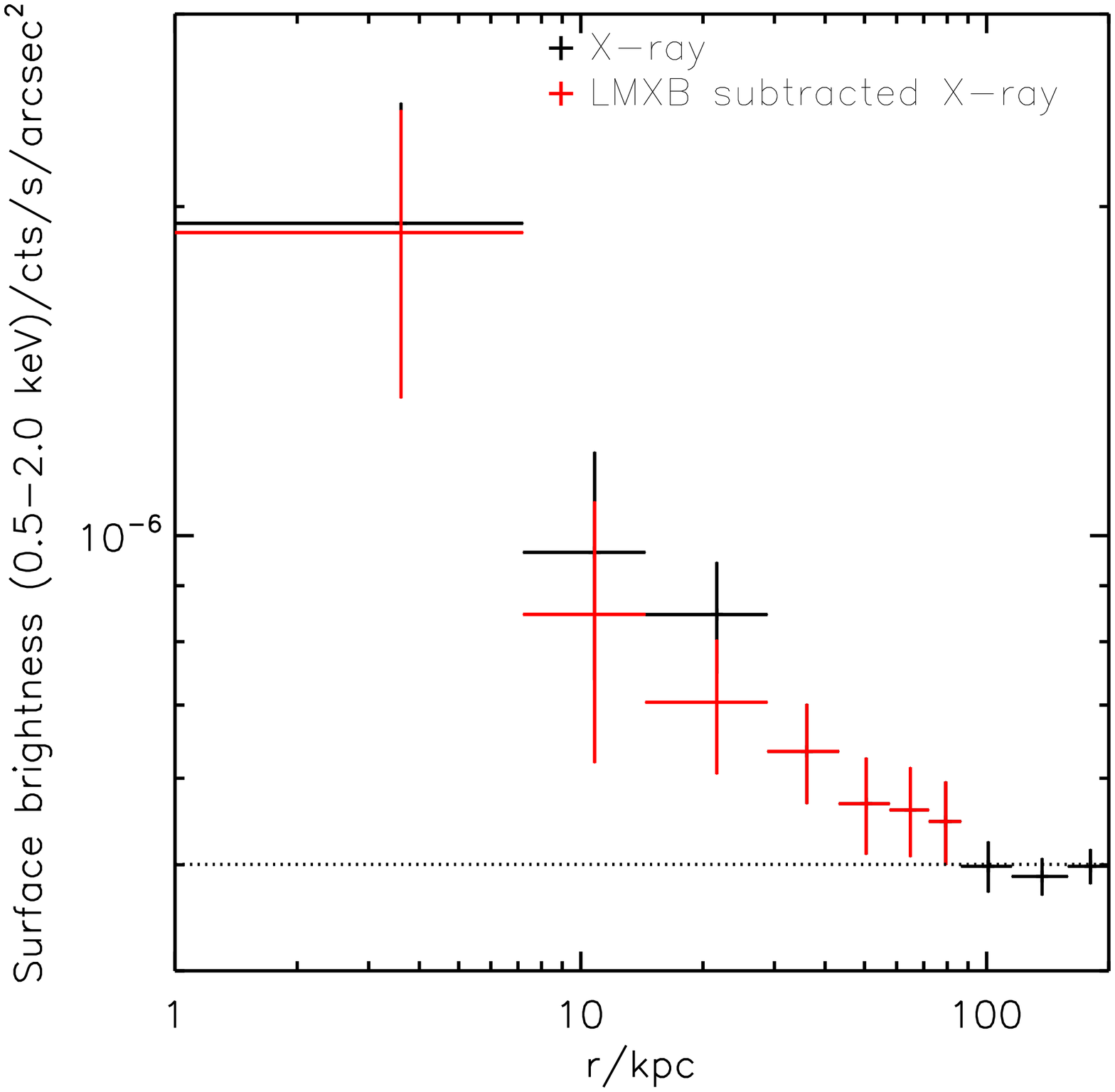,
        width=0.95\linewidth}
       \epsfig{figure=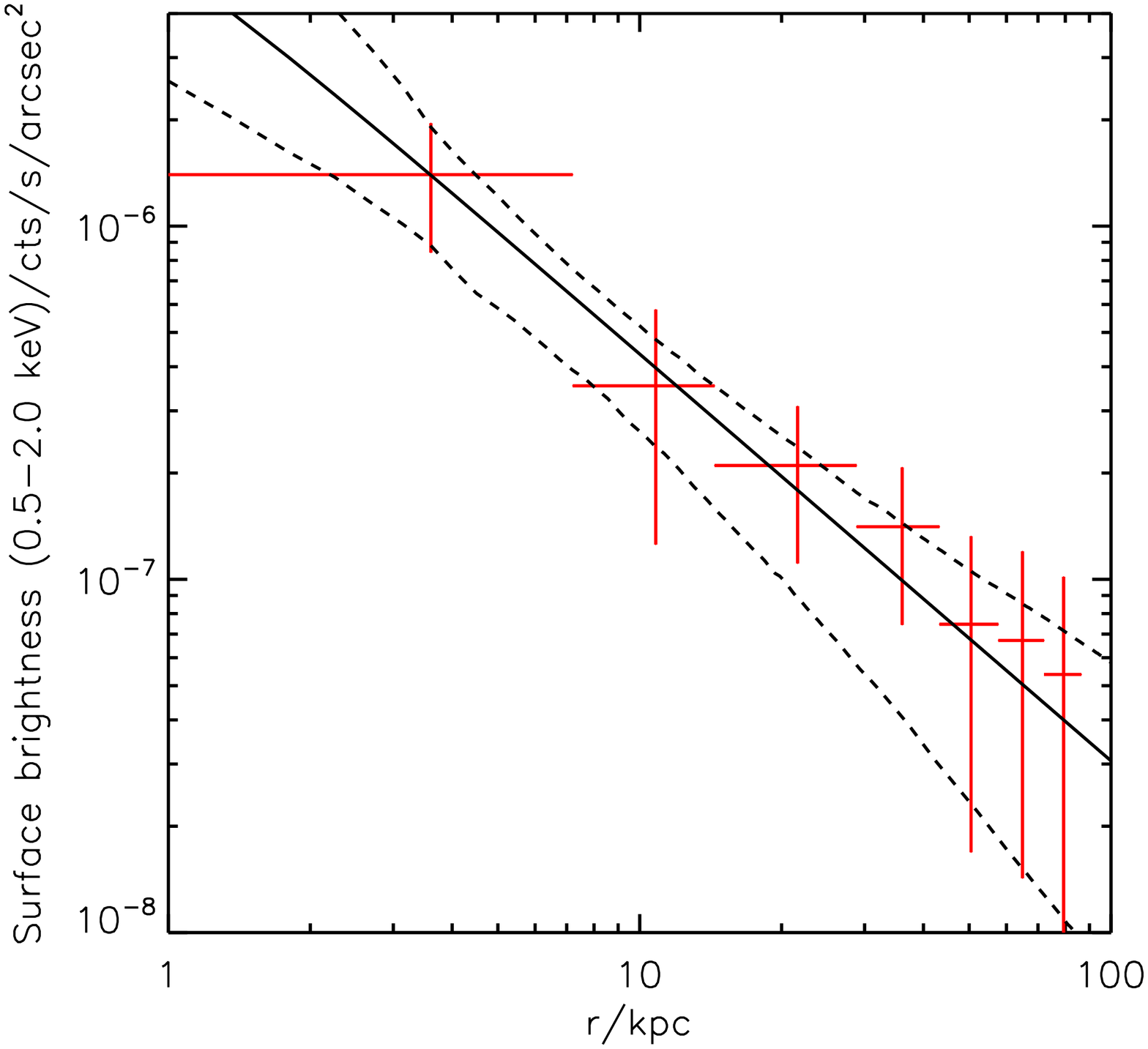,
        width=0.95\linewidth}       
        
               }
    
      \caption{\emph{Top}:Black points show exposure corrected surface brightness profile in the 0.5-2.0 keV band for the extended emission, avoiding the regions to the north and south of the galaxy where the jets may have removed gas from the halo. Red points show the remaining signal from the hot halo when the expected contribution from LMXBs is removed. In the 30-80 kpc range the LMXB contribution is negligible so the red points overlap with the black points exactly. \emph{Bottom}: The best fit profile to the exposure corrected, background subtracted and LMXB subtracted surface brightness profile. The systematic uncertainties in the background level have been propagated through into the error bars shown. The dashed lines show the 1-$\sigma$ range of the fitting from the Monte Carlo analysis performed.   }
      \label{SBprofile}
  \end{center}
\end{figure}

In Fig. \ref{SBprofile} (top panel) we show the soft band (0.5-2.0 keV) surface brightness profile over the halo as the black points, avoiding the regions to the north and south of the galaxy where the jets are located and where there appears to be little X-ray emission. The background level is shown as the dotted horizontal line. The emission is detected out to a radius of around 80 kpc, similar to the extent found for the halo around UGC 12591 in \citet{Dai2012}.

We need to account for contamination from LMXBs, whose X-ray emission also declines with radius from the centre of the galaxy. To do this, we follow \citet{Anderson2011} and find the surface brightness profile for the same regions in a hard band, 2.0-6.0 keV, in which the emission should be dominated by LMXB emission due to the low temperature of the expected hot halo (kT $\sim$ 0.6 keV). We then calculate the expected emission in the 0.5-2.0keV band from these LMXBs, assuming them to have a powerlaw spectrum with an index of $\Gamma$=1.56. This is motivated by \citet{Irwin2003}, which found a universal spectrum for the integrated emission from low mass X-ray binaries of this form. We calculate using the ACIS-S response files that for each count in the 2.0-6.0 keV band from LMXBs, we expect there to be 2 counts in the 0.5-2.0keV band. 

We subtract the calculated LMXB contamination from the soft band surface brightness profile, and this result is shown by the red points in Fig. \ref{SBprofile} (top panel). The LMXB contamination constitutes only around 10 percent of the total emission in the soft band, so the effect is not substantial. 

As a further check, we compared our estimated contribution from LMXBs with the contribution expected based upon the average LMXB luminosity function and the galaxy stellar mass, and found these to agree well. 

\begin{figure}
  \begin{center}
    \leavevmode
    \hbox{
      \epsfig{figure=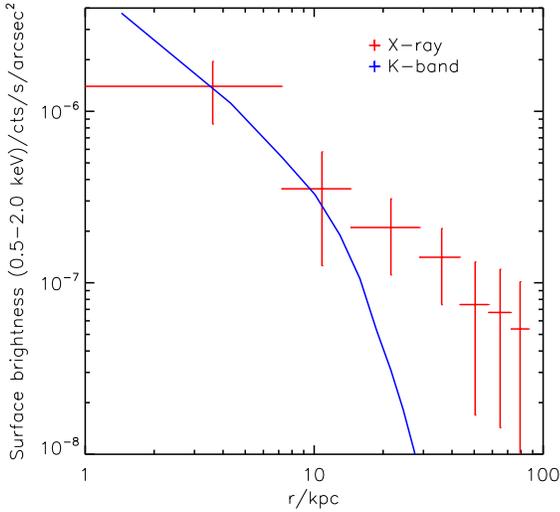,
        width=0.95\linewidth}
               }
    
      \caption{Plot comparing the LMXB corrected, exposure and background subtracted X-ray surface brightness profile from the bottom panel of Fig. \ref{SBprofile} (red points) with the background subtracted 2MASS K-band surface brightness profile (blue line). The K-band profile has been scaled so that it matches the innermost bins of the X-ray profile. We see that the X-ray emission is clearly more extended than the K-band emission, reaching out to at least 80 kpc from the galaxy centre.}
      \label{compare_Kband}
  \end{center}
\end{figure}

In Fig. \ref{compare_Kband} we compare the shape of the background subtracted (and LMXB corrected) X-ray surface brightness profile with that of the background subtracted K-band profile from 2MASS data. The 2MASS profile has been scaled to match the X-ray profile in the innermost bins. We see that the X-ray surface brightness profile is clearly much more extended than the K-band profile, with X-ray emission out to at least 80 kpc from the galaxy centre.

\section{Halo mass composition}

%
%

\subsection{Stellar mass}
\label{stellarmasssec}

In the central regions of the galaxy, the mass is dominated by the stellar mass, which can be estimated using the total K band absolute magnitude of -26.15 determined from 2MASS data. Using a mass to light ratio of 0.78 with a range 0.6-0.95 (\citealt{Bell2003}, \citealt{Dai2012}), this yields a stellar mass of 4.6$^{+1.0}_{-1.0}$ $\times$ 10$^{11}$ M$_{\odot}$.

\subsection{Star formation and supernova rate}
\label{starformationrate}

To estimate the star formation rate of the galaxy, we use the far ultraviolet (FUV)
flux obtained from GALEX, which is 78.06 $\mu$Jy at an effective wavelength of 1539 \AA. The FUV luminosity is thus 1.1$\times 10^{28}$ erg s$^{-1}$ Hz$^{-1}$. Using equation 2 from \citet{Bell2001}, this yields a star formation rate of 1.6 M$_{\odot}$ yr$^{-1}$, which is just slightly higher than the star formation rate found in normal galaxies.  

To determine whether some fraction of the gas has been uplifted from the disk by supernova (SN) driven winds, we need to calculate the SN rate per unit surface area (the specific SN rate). \citet{Strickland2004} estimate that SN blowout should theoretically occur when the specific SN rate exceeds 25 SN Myr$^{-1}$ kpc$^{-2}$. To estimate the expected supernova rate, we assume a Salpeter initial mass function ($N(M) \propto M^{-7/3}$), and calculate that the fraction of stellar mass in stars massive enough to produce supernovae (those between 8M$_{\odot}$ and 20 M$_{\odot}$), $f_{M,SN}$ is:
\begin{eqnarray}
f_{M,SN} = \frac{\int_{8}^{20}M \times M^{-7/3} dM}{\int_{0.1}^{20}M \times M^{-7/3} dM} = 7.2 \,\,  \rm{percent}
\end{eqnarray}
so of the 1.6 M$_{\odot}$ yr$^{-1}$ of star formation, we expect $0.072\times1.6=0.12$ M$_{\odot}$ yr$^{-1}$ to go into supernovae. The median mass of a star going supernova assuming the Salpeter IMF, $\overline{M_{SN}}$, is given by:
\begin{eqnarray}
\frac{\int_{8}^{\overline{M_{SN}}}M \times M^{-7/3} dM}{\int_{0.1}^{20}M \times M^{-7/3} dM} = 1/2
\end{eqnarray}
from which we obtain $\overline{M_{SN}}$=12.2 M$_{\odot}$. So the expected supernova rate is 0.12 M$_{\odot}$ yr$^{-1}$/12.2 M$_{\odot}$ = 0.01 SN yr$^{-1}$. Taking the major axis diameter of 22.6 arcsec = 33 kpc from NED\footnote{https://ned.ipac.caltech.edu/}, the specific SN rate is 11 SN Myr$^{-1}$ kpc$^{-2}$, which is lower than the blowout limit of 25 SN Myr$^{-1}$ kpc$^{-2}$. So we conclude that gas uplifted by SN driven winds has not contributed significantly to the hot halo we observe in X-rays.

\subsection{Hot gas mass}
\label{halomasssec}

The following extrapolations assume that the gas is in hydrostatic equilibrium in the potential well, and are provided as an indication of what we obtain for the gas mass if hydrostatic equilibrium is assumed. Since it appears that the AGN has significantly disrupted the gas, it is possible that the gas is not in hydrostatic equilibrium, and so these results should be interpreted with this caveat. The effect of the AGN feedback on the hot halo gas makes it challenging to relate the observed gas mass with simulations of galaxy formation.

Here we calculate the hot halo mass out to 80 kpc ($\simeq$0.15r$_{200}$), the maximum radius out to which we detect an X-ray signal, excluding the central 0.05$r_{200}$=15.5kpc region of the galaxy. The hot halo appears asymmetric, and gas may have been removed along the jet directions. To account for this possibility, we calculate two different estimates of the hot halo mass.  The first is the total hot gas mass of an original halo which we assume to be spherical, but then the jets have removed gas perpendicular to the disk. This provides a realistic upper limit to the total hot gas mass that the system could have had originally before any action of the jets. This uses the surface brightness profile of the galaxy, excluding the jet directions. We then work out the total mass assuming spherical symmetry, in effect reconstructing the gas mass which we assume to have been removed to the north and south. 

To do this, we fit the LMXB and background subtracted soft band surface brightness profile presented in the bottom panel of Fig. \ref{SBprofile} with a beta model:

\begin{eqnarray}
S(r) = S_{0}\left[ 1+ \left( \frac{r}{r_{0}}\right)^2  \right] ^{0.5-3 \beta} 
\label{SB_betamodel}
\end{eqnarray}
The best fit is shown as the solid curve in the bottom panel of Fig. \ref{SBprofile}, which has best fitting parameters of $S_{0}=1.4^{+0.3}_{-0.2}  \times10^{-5}$ cts/s/arcsec$^{2}$, $r_{0}=0.5^{+0.2}_{-0.1}$ kpc and $\beta=0.36^{+0.04}_{-0.03}$. 
 
These 1-$\sigma$ errors were obtained by performing a Monte Carlo method with 100,000 trials, and the 1-$\sigma$ range of the fits is shown by the dashed lines in the bottom panel of Fig. \ref{SBprofile}.
 
The beta model is the theoretical expectation for an isothermal sphere in hydrostatic equilibrium, and has been found to provide reasonable fits to hot gas halos around elliptical galaxies, and to the large scale intracluster medium emission in galaxy clusters and groups. Under the assumption of constant metallicity and uniform temperature, the gas density profile is then given by:

\begin{eqnarray}
n(r) = n_{0}\left[ 1+ \left( \frac{r}{r_{0}}\right)^2  \right] ^{-1.5\beta} 
\label{n_betamodel}
\end{eqnarray}

Using the ACIS-S response files, we can convert the observed surface brightness profile into the normalisation of an absorbed \textsc{apec} component at each radius. We fix the \textsc{apec} parameters to typically expected values of kT=0.6 keV and Z=0.1Z$_{\odot}$ (\citealt{Bogdan2013a}). We use the abundance tables of \citet{Grevesse1998}. The redshift is fixed to z=0.0755, and we fix the absorbing column to the LAB survey (\citealt{LABsurvey}) value of $3\times10^{20}$ cm$^{-2}$. Unfortunately, due to the low count rate, we are unable to determine these parameters accurately by direct spectral fitting. We can then find the density profile, which has the beta model form of equation \ref{n_betamodel} with $n_{0}=0.08^{+0.03}_{-0.02}$ cm$^{-3}$. 

Integrating this density profile from 15.5kpc to 80 kpc (0.05 to $\simeq$0.15$r_{200}$) and assuming spherical symmetry to reconstruct the mass assumed to have been lost along the jet directions, we obtain a hot gas mass of $M_{gas}(0.05<r<0.15r_{200})$ $2^{+1}_{-1}\times10^{10}$ M$_{\odot}$, assuming a temperature of 0.6 keV and a metallicity of 0.1 Z$_{\odot}$. The results are not significantly affected by excluding the central two bins when fitting the surface brightness profile. The hot gas mass determination depends significantly on the actual metal abundance and temperature. A higher abundance leads to more emission from the Fe L line, giving a lower gas density for a fixed count rate. Taking the range 0.0-1.0 Z$_{\odot}$ for the metallicity, and the range 0.2-1.0 keV for kT, the corresponding possible range for the hot gas mass $M_{gas}(0.05<r<0.15r_{200})$ is 0.7-4.0$\times 10^{10}$ M$_{\odot}$.

In our second hot halo mass estimate, we just find the mass of hot gas currently in the system, taking into account the possibility that the jets have removed the gas they have interacted with to the north and south. We fit a beta model to the surface brightness profile and proceed as before, but now exclude cones of opening angle 90 degrees to the north and south when performing the integration, which is a upper limit to the amount of mass removed. We find a hot gas mass $M_{gas}(0.05<r<0.15r_{200})$ of $1.2^{+2.0}_{-0.7}\times10^{10}$ M$_{\odot}$, where these errors factor in the systematic uncertainty in the metal abundance and gas temperature as before.

\subsection{The Energetics of the AGN}
\label{energetics}

Here we explore the energetics of whether it is feasible that the AGN could have removed gas from the potential well of the galaxy, or if it is only powerful enough to move gas to larger radii. \citet{Cavagnolo2010} have explored the relationship between the synchrotron radio luminosity and the jet power, which we can use to estimate the feedback power due to the AGN in J2345-0449. 

Given the 1.4GHz radio luminosity of the lobes of 2.5($\pm$0.3)$\times 10^{31}$ erg s$^{-1}$ Hz$^{-1}$ obtained in \citet{Bagchi2014} (so a radio power of 3.5$\times 10^{40}$ erg s$^{-1}$), we use the following best fit relation from \citet{Cavagnolo2010} to obtain the jet power:
\begin{eqnarray}
\log P_{jet} = 0.75 (\pm 0.14) \log P_{1.4} +1.91 (\pm 0.18)
\label{Cavagnolorelation}
\end{eqnarray}
where $P_{jet}$ is in units of 10$^{42}$ erg s$^{-1}$, and the 1.4GHz radio power $P_{1.4}$ is in units of 10$^{40}$ erg s$^{-1}$. This yields a jet power of 2$\times 10^{44}$ erg s$^{-1}$. However, as can be seen in figure 1 of \citet{Cavagnolo2010}, the scatter in the scaling relation between the jet power and the radio power is large, at least an order of magnitude at the radio power being considered. This large scatter prevents us from performing a highly accurate calculation, limiting us to order of magnitude estimates. 

Given that the distance from the AGN to the outer part of the outer radio lobe is 780 kpc=2.5 million light years, the absolute minimum time the jets could have been active for is 2.5 Myr. The time for which jets are active is typically in the range 10-100 Myr. If we assume 10 Myr, the total feedback energy deposited into the hot halo given a jet power of 2$\times 10^{44}$ erg s$^{-1}$ is 7$\times10^{58}$ erg.

If we assume that the halo was originally spherically symmetrical, and that gas has been removed from the north and south in the 20-80 kpc range, we can use our density profiles obtained earlier in section \ref{halomasssec}
to estimate that the mass of gas lying in this 20-80 kpc range is $\sim 2\times10^{10}$ M$_{\odot}$. 

To calculate the depth of the gravitational potential in this radial range, we assume that the dark matter halo has an NFW profile, so that the potential in the dark matter halo,$\Phi_{DM}$, is given by (\citealt{Lokas2001}):
\begin{equation}    \label{c9}
    \frac{\Phi(s)_{DM}}{V_{200}^2} = - \frac{\ln (1 + c s)}{s(\ln (1+c) - c/(1+c))} \ ,
\end{equation}
where $s=r/r_{200}$, the concentration parameter $c$ is set to 10 (as typically found for haloes with virial masses around $10^{13}$ M$_{\odot}$ in numerical $\Lambda$CDM simulations, e.g. \citealt{Bullock2001}), and $V_{200}$ is the circular velocity at $r=r_{200}$ given by:
\begin{equation}    \label{c10}
    V_{200}^2 = \frac{4}{3} \,\pi \,G \,r_{200}^2 \,200 \,\rho_c^0 \ .
\end{equation}
where $\rho_c^0$ the critical density of the universe. 

Given that $r_{200}$=450kpc, we find that in the radial range, 20-80kpc, the average gravitational binding energy per unit mass due to the dark matter halo alone is 5$\times10^{48}$ erg/M$_{\odot}$. When we factor in the baryonic mass of the galaxy, the total average gravitational binding energy in the radial range 20-80 kpc rises slightly to 5.6$\times10^{48}$ erg/M$_{\odot}$. This means that on average it takes 5.6$\times10^{48}$ erg of energy to move one solar mass of gas from the 20-80 kpc radial range to infinity (i.e, no longer gravitationally bound to the galaxy). 

Using our estimate of 7$\times10^{58}$ erg for the total amount of feedback energy deposited into the hot halo, the total mass that could be removed from the galaxy is of the order 7$\times10^{58}$/ 5.6$\times10^{48}$ $\sim$ $10^{10}$ M$_{\odot}$. This is of the same order of magnitude as the total amount of gas lying in the 20-80kpc range ($2\times10^{10}$ M$_{\odot}$) we estimated earlier assuming that the hot halo was originally spherically symmetric.

We therefore conclude that it is energetically possible that the AGN has completely removed the gas to the north and south in the 20-80 kpc range from the galaxy. However, given the order of magnitude of scatter in the radio power to jet power scaling relation, and the uncertainty in the life span of the jets, it is not possible to make a more precise estimate of the amount of gas that could have been removed or redistributed to higher radii.

It is possible that the heating from the AGN has raised the temperature of the gas and increased its detectability. Due to the uncertainties in how much energy has been deposited in the hot halo from the AGN, and the way this heating has been distributed, it is not possible to estimate the magnitude of such a temperature increase.

\section{Black hole mass}

\citet{Bagchi2014} found that the stellar velocity dispersion within the central 2.35 kpc region of the galaxy is exceptionally large ($\sigma=379\pm25$ km s$^{-1}$ along the major axis, and $351\pm25$ km s$^{-1}$ along the minor axis), strongly suggesting the presence of a highly massive black hole. \citet{Bagchi2014} also found that the galaxy has a pseudobulge with a dynamical mass of 1.07($\pm$0.14) $\times10^{11}$ M$_{\odot}$.

\citet{Bagchi2014} found that, when using the black hole mass to bulge mass correlation of \citet{Marconi2003}, the black hole mass is  2.5$\pm0.5\times10^{8}$ M$_{\odot}$. However \citet{Bagchi2014} also find that a much larger mass of  1.4$\times10^{9}$ M$_{\odot}$ can be found using the $M_{\rm BH} - \sigma$ relation of \citet{Gultekin2009}, using the velocity dispersion averaged within the central 2.35 kpc region along the minor axis, which they find to be 351($\pm$25) km s$^{-1}$. As the galaxy lacks a classical bulge (it has a pseudobulge), it is challenging to accurately estimate the black hole mass using scaling relations with the bulge mass, which may cause the black hole mass estimate using this method to be underestimated.

Using the more recent $M-\sigma$ relation from \citet{McConnell2013}, we find using $\sigma$=351
 km s$^{-1}$ as before that we obtain an even higher black hole mass of $5\times 10^{9}$ M$_{\odot}$.
 
We can now use the X-ray and radio luminosities of the central AGN to obtain another estimate the central black hole mass using the fundamental plane of \citet{Gultekin2009}:
\begin{eqnarray}
\log M_{\rm BH,8} = \mu_{\rm 0} + c_{\rm r} \log L_{\rm R,38} +c_{\rm x} \log L_{\rm X,40}
\label{fundamentalplane}
\end{eqnarray}
where $M_{\rm BH,8}$ is the black hole mass in units of $10^{8}$ M$_{\odot}$, $L_{\rm R,38}$ is the 5GHz radio luminosity in units of $10^{38}$ erg s$^{-1}$, and $L_{\rm X,40}$ is the 2-10keV X-ray luminosity in units of $10^{40}$ erg s$^{-1}$. The constants are $\mu_{\rm 0}=0.19\pm0.19$, $c_{\rm r}=0.48\pm0.16$ and $c_{\rm x}=-0.24\pm0.15$. From our Chandra data we obtain $L_{\rm X,40}$=9.9$\pm1.5$ erg s$^{-1}$. This X-ray luminosity was obtained by using the Chandra response files and an assumed spectral shape of a powerlaw of index 1.9, as commonly found for AGN, to convert the observed X-ray count rate of the AGN in the 0.5-2.0 keV band into a 2-10keV flux and then into the luminosity in this band. The 5GHz radio luminosity obtained in \citet{Bagchi2014} is 
$L_{\rm R,38}$=45$\pm5$ erg s$^{-1}$, using VLA data. This yields a black hole mass of 5$\times10^{8}$ M$_{\odot}$, though there is considerable intrinsic scatter in the mass direction of this fundamental plane relation of 0.77 dex. The range of black hole mass taking into account the scatter is $1\times10^{8}$ - $3\times10^{9}$ M$_{\odot}$, which encompasses the possibility that the central black hole is exceptionally massive (i.e. $>10^{9}$ M$_{\odot}$).

One main implication of an extremely massive black hole in this galaxy is that significant amounts of AGN feedback energy are possible, which can disrupt the gas in the hot halo around the galaxy.

\section{Conclusions}  

Using deep Chandra observations we detect an extended X-ray halo around the massive spiral galaxy J2345-0449, reaching out to 80 kpc from the galaxy centre, far in excess of the the extent of the optical emission. The X-ray halo appears elongated along the plane of the galaxy, which is possibly the result of the powerful jets removing gas above and below the galactic plane. It is possible that the rotation of the halo may also contribute to its flattened appearance in X-rays. We also detect extended X-ray emission which is spatially coincident with the inner and outer radio lobes, and may be the result of inverse Compton scattering.

In J2345-0449, we may be seeing first hand the way AGN feedback expels gas from massive spiral galaxies, contributing to at least some of the baryon deficit below the mean baryon fraction typically seen in the observations of hot halos around massive spiral galaxies. We have calculated that it is energetically feasible that the AGN feedback operating in this system could have removed a significant fraction of the gas within the 20-80 kpc region of the hot halo we have observed. 

Using the X-ray and radio luminosities of the central AGN, the fundamental plane of \citet{Gultekin2009} gives a central black hole mass of 5$\times10^{8}$ M$_{\odot}$, with a range of $1\times10^{8}$-$3\times10^{9}$ M$_{\odot}$ when the scatter in the plane is taken into account. This method of calculating the black hole mass produces results which are consistent with the possibility of an exceptionally massive ($>$10$^{9}$ M$_{\odot}$), as suggested by the $M-\sigma$ relations of \citet{Gultekin2009} and \citet{McConnell2013}. However, due to the large observational scatter in the fundamental plane relation, a tighter constraint cannot be made.

\section*{Acknowledgements}

SAW and ACF acknowledge support from ERC Advanced
Grant FEEDBACK. This
work is based on observations obtained with the \emph{Chandra} observatory, a NASA mission.

\bibliographystyle{mn2e}
\bibliography{Galaxy_halo_paper}

\appendix
\section[]{}
\label{sec:appendix}
      \label{AppendixA}

\begin{figure}
  \begin{center}
    \leavevmode
    \hbox{
      \epsfig{figure=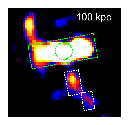,
        width=0.95\linewidth}
               }
    
      \caption{The extraction region used to obtain the counts in the hot halo is shown as the dashed green box, excluding the central 0.05$r_{200}$ region shown as the dashed green circle. The extraction region for the inner southern lobe is shown by the dashed white boxes.}
      \label{haloregion}
  \end{center}
\end{figure}

%
%
%
%
%
%
%


\end{document}